\documentclass{svjour2}                    % onecolumn
\smartqed  % flush right qed marks, e.g. at end of proof
\usepackage{graphicx}
\begin{document}

\title{Current Sensing Noise Thermometry: A fast practical solution to low temperature measurement%\thanks{Grants or other notes
}


\author{A. Casey         \and
        F. Arnold         \and		
				L. V. Levitin        \and
				C. P. Lusher         \and
				J. Saunders					\and
				A. Shibahara         \and
				H. van der Vliet	\and
				D. Drung         \and
				Th. Schurig        \and
				G. Batey         \and
				M. N. Cuthbert         \and
				A. J. Matthews
}


\institute{A. Casey, F. Arnold, L. V. Levitin, C.P. Lusher, J. Saunders, A. Shibahara, H. van der Vliet \at
              Department of Physics, Royal Holloway University of London, Egham, Surrey, TW20 0EX, United Kingdom \\
              Tel.: +44 (0)1784 414351\\
              Fax: +44 (0)1784 472794\\
              \email{a.casey@rhul.ac.uk}           %  \\
           \and
           D. Drung, Th. Schurig \at
              Physikalisch-Technische Bundesanstalt, Abbestrasse 2-12, D-10587 Berlin, Germany\\
						 \and
           G. Batey, M. N. Cuthbert, A. J. Matthews \at
              Oxford Instruments Omicron Nanoscience, Tubney Woods, Oxon, OX13 5QX, United Kingdom\\
						\email{Anthony.Matthews@oxinst.com}
}

\date{Received: date / Accepted: date}

\maketitle

\begin{abstract}
We describe the design and performance of a series of fast, precise current sensing noise thermometers. The thermometers have been fabricated with a range of resistances from  1.290~$\mathrm{\Omega}$ down to 0.2~m$\mathrm{\Omega}$. This results in either a thermometer that has been optimised for speed, taking advantage of the improvements in superconducting quantum interference device (SQUID) noise and bandwidth, or a thermometer optimised for ultra-low temperature measurement, minimising the system noise temperature. With a single temperature calibration point, we show that noise thermometers can be used for accurate measurements over a wide range of temperatures below 4~K. Comparisons with a melting curve thermometer, a calibrated germanium thermometer and a pulsed platinum nuclear magnetic resonance thermometer are presented. For the 1.290~$\mathrm{\Omega}$ resistance we measure a 1~\% precision in just 100~ms, and have shown this to be independent of temperature. 
\end{abstract}

\section{Introduction}
\label{intro}
There is a rapidly expanding use of low temperature platforms ($T < 1$~K), for fundamental science, sensitive instrumentation, and new technologies of potentially significant commercial impact. The availability of commercial ``push button'' cryogen-free dilution refrigerators is a key game-changing recent development, in which European companies have played the leading role. The availability of cryogen-free commercial platforms to 1~mK is a future prospect \cite{Casey2013}.

Precise measurement of the thermodynamic temperature of these low temperature platforms is crucial for their operation. However, despite the sophisticated nature of the cooling technology, reliable and convenient thermometers are unavailable.

In the temperature range below 1~K the dissemination of the kelvin is usually realised using secondary thermometers and superconducting reference points \cite{LounasmaaB,EnssB,PobellB}. Examples include: resistance thermometers; paramagnetic susceptibility; vibrating wire/tuning fork viscometry. These thermometers do not reliably cover the temperature range 1~K to 10~mK (a fortiori to 1~mK). The calibration of these thermometers against the PLTS-2000 \cite{PLTS2000,Rusby2002} is impractically time consuming and expensive. Resistance thermometers are intrinsically subject to shifts in calibration and prone to self heating and pick-up of electromagnetic interference. Paramagnetic-salt susceptibility thermometers are subject to degradation with time, and viscometric methods measure temperature inside the mixing chamber and not on the typical sample mounting platform.
 
The dependence of $^{3}$He melting pressure on temperature has been used as the laboratory standard in specialist laboratories for many years. The PLTS-2000 scale is based on the melting pressure of $^{3}$He but this is viewed as difficult to implement for the rapidly expanding community of users of ultra-low temperature platforms. $^{60}$Co nuclear orientation thermometry is a primary technique that relies on the known anisotropy of the decay of radioactive nuclei as a function of temperature but is only effective in a narrow temperature range, and is slow and user unfriendly. Nuclear magnetic susceptibility thermometry is likewise insensitive above 20~mK, and user unfriendly. Reliable and traceable thermometry below 1~K thus offers demanding challenges even to the specialist, and, despite the sophisticated low temperature platforms now readily available with ``push-button'' operation, a solution which is reliable, traceable and user-friendly has yet to be developed. 

A quote attributed to Prof O V Lounasmaa, founder of the Low Temperature Laboratory at Helsinki University of Technology (now Aalto University) in the context of the T \textless 1 K regime reads ``A person with one thermometer knows what the temperature is; a person with two will forever have doubt''. The challenge is to rectify this state of affairs.

The need to have a thermometer that measures thermodynamic temperature outside of international standards laboratories has led to the development of thermometers with a simple relationship between the measured parameter and the Boltzmann constant, $k_{\mathrm{B}}$, for example thermometers that couple noise to a SQUID either directly \cite{Lusher2001} or indirectly via the magnetic Johnson noise \cite{Varpula1993,Beyer2007,Engert2009,Enss2013} and Coulomb Blockade thermometers \cite{Pekola1994} based on electric conductance characteristics of tunnel junction arrays.

In this paper we describe recent advances in current sensing noise thermometer, exploiting low noise and high bandwidth dc SQUID amplifiers, for the high precision and rapid measurement of absolute temperature, covering the entire temperature range of interest from 4K to below 100 microK.

\section{Current Sensing Noise Thermometry}
\label{sec:1a}

The current sensing noise thermometer was pioneered by Giffard et al.\cite{Giffard1972,Webb1973} and developed by Webb and Washburn \cite{Webb1983}. The principle is to measure the thermal Johnson noise \cite{Johnson1928} in a resistor to determine its temperature. The performance of the thermometers developed in \cite{Giffard1972,Webb1973,Webb1983} was limited by the noise of the RF SQUID amplifiers; in \cite{Webb1983} the typical device noise temperature was a few mK. Using a low transition temperature dc SQUID as the front end amplifier the current sensing technique becomes a practical approach for measuring absolute temperature from 4.2 K down to sub mK temperatures, as shown in \cite{Lusher2001}; this work used commercial dc SQUIDs. The improved sensitivity of dc SQUIDs fabricated at PTB \cite{Drung2007} allows the use of sensor resistors with much larger resistance, thereby increasing the speed of the thermometer. In this paper we demonstrate experimentally an improvement in speed by orders of magnitude and we demonstrate that the percentage precision is independent of temperature.

The principle of operation of a current sensing noise thermometer is discussed in detail in our earlier paper \cite{Lusher2001} but is summarised below. The mean square open circuit noise voltage per unit bandwidth across the terminals of a resistor, $R$, at temperature $T$ is given by the Nyquist expression:
\begin{equation}
\label{eq:1}
\left\langle V^{2}_{N}\right\rangle=4k_{\mathrm{B}}TR
\end{equation}
This is a white noise spectrum over the frequency range of interest. In our current sensing noise thermometer, the resistor whose absolute temperature is to be measured is connected in series with the superconducting input coil of the readout SQUID through a superconducting twisted pair. The mean square noise current flowing in the SQUID input coil per unit bandwidth arising from the thermal noise in the resistor is then given by:
\begin{equation}
\label{eq:2}
\left\langle I^{2}_{N}\right\rangle=\frac{4k_{\mathrm{B}}T}{R}\left(\frac{1}{1+\omega^{2}\tau^{2}}\right)
\end{equation}
where $\omega = 2 \pi f$. The time constant $\tau=L/R$, where the total inductance in the input circuit $L$ includes the SQUID input coil inductance $L_{i}$ and any additional stray inductance including that of the twisted pairs. By increasing the resistance, it is possible to increase the bandwidth of the noise spectrum. However, this results in a reduced noise power from the resistor at $\omega \tau < 1$, which must be kept much greater than the SQUID noise in order to ensure a reasonable device noise temperature, $T\mathrm{_{N}}$, much less than the minimum temperature we wish to measure . The SQUID is operated in flux-locked loop (FLL) mode where the gain depends only on fixed, known parameters. Thus, if the value of the resistive sensor and the SQUID gain can be measured, a current noise spectrum can be used to determine absolute temperature using Equation \ref{eq:2}.
\section{Experimental Details}
\label{sec:1}
\begin{figure}
  \includegraphics{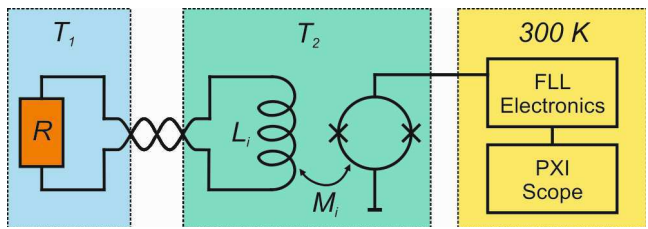}
\caption{Schematic diagram of the current sensing noise thermometer. The dc SQUID amplifies the thermal noise current of the resistor with the output captured by a digitiser. $T_{1}$ can be at a different temperature to $T_{2}$, the temperature of the SQUID.}
\label{fig:1}       % Give a unique label
\end{figure}
Figure \ref{fig:1} shows a schematic of the measurement set-up. The SQUID sensor used in this work is a PTB two-stage SQUID \cite{Drung2007} with an input coil inductance of $L_{i}=1.05$~$\mu$H and a mutual inductance $M_{i}$ between the SQUID and the input coil equivalent to $1/M_{i}=0.29$~$\mathrm{\mu}$A/$\phi_{0}$, where $\phi_{0}$ is the flux quantum. When operated with a 10~k$\mathrm{\Omega}$ feedback resistor, the FLL gain is 0.42~V/$\phi_{0}$. The SQUID is located remotely from the sensor resistor, allowing the resistor to be placed in environments hostile to SQUID operation. In this case the SQUID is mounted on the continuous heat-exchanger plate of the dilution fridge at approximately 100~mK. At this temperature, the white flux noise floor of the SQUID is $0.58~\mathrm{\mu}\phi_{0}$Hz$^{-1/2}$, equivalent to a coupled energy sensitivity, $\epsilon_{c}=22 h$ in units of Planck's constant, where:
\begin{equation}
\label{eq:3}
\epsilon_{c}=\frac{1}{2}L_{i}\frac{\left\langle \phi^{2}_{N}\right\rangle}{M^{2}_{i}}
\end{equation}
and  $\left\langle \phi^{2}_{N}\right\rangle$ is the flux noise power. This can be compared to  $\epsilon_{c}=500 h$ for the commercial Quantum Design \cite{QuantDesign} SQUID operated at 4.2~K used in our previous work \cite{Lusher2001,Lusher2003}. This lower noise floor corresponds to a lower amplifier noise temperature  $T\mathrm{_{N}}$. Thus through choice of resistor the user can minimise the noise temperature for absolute precision or increase the bandwidth of the measurement for increased speed. The primary focus of this work was to establish the optimum choice of resistance value for operation of the thermometers on a dilution refrigerator (with a typical base temperature of 5~mK). In addition in order to investigate the $^{3}$He melting curve fixed points on the provisional low temperature scale PLTS-2000 we need a thermometer that can be reliably operated below 1~mK. For these reason we chose sensor resistance values of 0.2~m$\mathrm{\Omega}$, 0.14~$\mathrm{\Omega}$ and 1.290~$\mathrm{\Omega}$.

\subsection{Construction of noise sensors}
\label{sec:2}
\begin{figure}
  \includegraphics[width=0.5\textwidth]{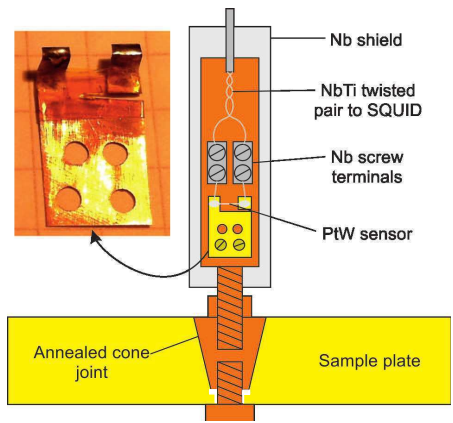}
\caption{Schematic of a PtW noise thermometer mounted onto a copper sample plate. The plate has either a thermal link to the mixing chamber of the dilution fridge or the sensor is mounted directly onto the copper nuclear demagnetisation stage.}
\label{fig:2}       % Give a unique label
\end{figure}
The 0.2~m$\mathrm{\Omega}$ resistor was fabricated with a copper foil as described in \cite{Lusher2001}. Using this material it is not practical to make a sensor with a resistance greater than $\approx$~10~m$\mathrm{\Omega}$. A higher resistivity material was needed for the 0.14~$\mathrm{\Omega}$ and 1.290~$\mathrm{\Omega}$ sensors. For this work 50~$\mu$m diameter platinum tungsten wire (Pt92\%/W8\%) \cite{Goodfellow} was used. PtW was chosen as it has a weakly temperature dependant resistivity as well as having a very small magneto-resistance \cite{Cieloszyk1975}. Figure \ref{fig:2} is a schematic diagram of the sensor resistor and the sample plate. The required length of PtW wire is achieved by hard soldering it to two pieces of 100~$\mu$m thick copper foil separated by a well-defined gap, using Johnson Matthey EasyFlo2 solder. One foil is electrically grounded and consequently heat sunk to a large annealed copper holder with a copper screw contact. The other foil is electrically isolated from the holder. The sensor resistor is then connected to a pair of niobium screw terminals via short sections of CuNi clad NbTi wire. The screw terminals are connected to the remote SQUID input coil through a shielded NbTi superconducting twisted pair. The resistor and copper former are enclosed in a niobium can for shielding and heat sunk to a copper sample plate with an M4 screw and cone joint as depicted in Figure \ref{fig:2}. For the tests of the  1.290~$\mathrm{\Omega}$ sensor the sample plate was mounted at the mixing chamber plate of the dilution fridge, the  0.2~m$\mathrm{\Omega}$ and 0.14~$\mathrm{\Omega}$ sensors were mounted directly on the top plate of a copper nuclear demagnetisation stage. 
\subsection{Characterisation of the sensors}
\label{sec:3}
\begin{figure}
  \includegraphics[width=0.75\textwidth]{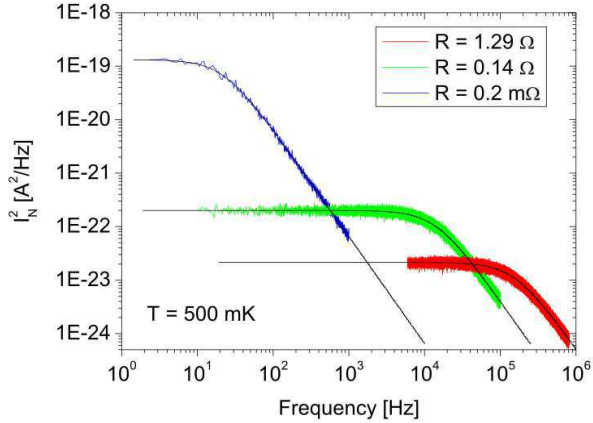}
\caption{Current noise spectra at 500~mK for the three noise sensors evaluated. Solid lines are fits using equation \ref{eq:2}.}
\label{fig:3}       % Give a unique label
\end{figure}
In principle, the current sensing noise thermometer is a primary thermometer; if the resistance of the sensor and the transfer function of the SQUID are known accurately. In practice however, it is more convenient to stabilise at a known reference temperature and determine the parameters $R$ and $L$ from a fit of the output noise power. For this single point calibration of our thermometers, the sample plate was held at 500~mK, measured by the melting curve thermometer. The temperature stabilisation was realised with a digital PID temperature controller to within 5~$\mu$K. The SQUID output voltage is read out using room temperature Magnicon XXF electronics \cite{Magnicon} and is captured with a PXI digitiser card \cite{NationalInstruments}. Time traces were captured with sampling frequencies of 50~kS/s, 500~kS/s, 2.5~MS/s for 0.2~m$\mathrm{\Omega}$, 0.14~$\mathrm{\Omega}$ and 1.290~$\mathrm{\Omega}$ sensors respectively, with a record length of 2$^{20}$ points. 1000 time domain traces were captured for each sensor.  Each captured time trace is individually Fourier transformed and the power spectral density is averaged in the frequency domain for improved precision. Before fitting, it was necessary to remove the noise power contribution of the SQUID, which was measured at high frequencies and assumed to be frequency independent over the range used for the fits.
The current noise spectra obtained in this way are shown in figure \ref{fig:3}. Fits were performed on the resulting current noise spectrum using Equation \ref{eq:2}, with $T=500$~mK. The resulting values for the resistance of the noise sensor $R$ and total inductance of the circuit $L$ are summarised in table \ref{tab:1}. These parameters were constrained in the fitting routine to determine the temperature of subsequent noise traces.

\begin{figure}
  \includegraphics[width=0.75\textwidth]{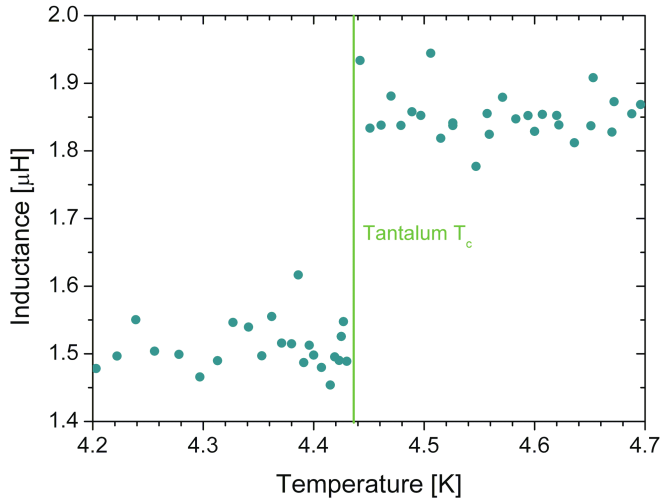}
\caption{Total inductance, $L$, of the noise thermometer input circuit as a function of temperature measured on the calibrated Germanium resistance thermometer. A sharp decrease in inductance is observed when the noise thermometer cools through the superconducting transition of the integrated tantalum fixed point device.}
\label{fig:4}       
\end{figure}

An alternative method to obtain the single fixed point calibration is to integrate a superconducting fixed point directly into the input circuit of the thermometer. An additional coil can be added to the superconducting flux transformer,wound around a sample with a known superconducting transition temperature; thermalized at the same point as the resistive sensor. The total inductance, $L$, of the input circuit is then sensitive to the state of the superconducting fixed point. Figure \ref{fig:4} shows the temperature dependence of the inductance measured in a current sensing noise thermometer in which we have included an in-situ tantalum \cite{Milne1961} fixed point device ( $T\mathrm{_{c}}$ = 4.45 K). Here 12 turns of NbTi wire are wound around a tantalum bobbin forming part of the superconducting flux transformer. The sharp transition of the tantalum is clearly visible in the noise thermometer measurement.
\begin{table}
\caption{Summary of the parameters of the noise thermometers evaluated,  $T\mathrm{_{N}}$ is measured from the high frequency noise.}
\label{tab:1}       % Give a unique label
\begin{tabular}{llll}
\hline\noalign{\smallskip}
Resistance & Total Inductance &  $T\mathrm{_{N}}$ & Bandwidth (3dB) \\
\noalign{\smallskip}\hline\noalign{\smallskip}
0.20436~m$\mathrm{\Omega}$  & 1.5016~$\mu$H & $<$1~$\mu$K & 22 Hz\\
0.13781~$\mathrm{\Omega}$ & 1.5483~$\mu$H  & 40~$\mu$K & 14.5 kHz\\
1.290~$\mathrm{\Omega}$ & 1.315~$\mu$H & 400~$\mu$K & 156 kHz \\
\noalign{\smallskip}\hline
\end{tabular}
\end{table}

\section{Performance}
\label{sec:4}
\begin{figure}
  \includegraphics[width=0.75\textwidth]{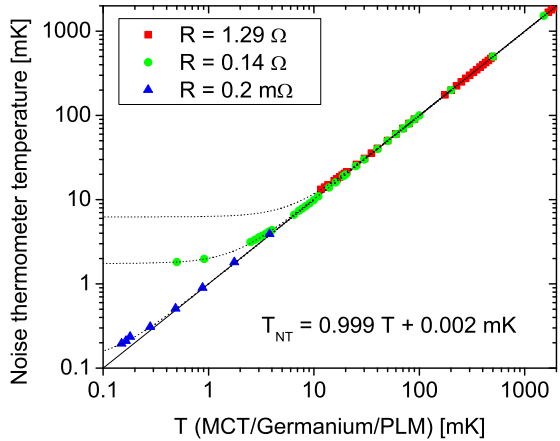}
\caption{Comparison of temperature obtained from the current sensing noise thermometers with that of a $^{3}$He melting curve thermometer, a calibrated germanium resistance thermometer and a pulsed platinum NMR thermometer over a wide temperature range.  The linear fit (solid line) to high temperature data is also shown. The decoupling of the thermometer at the lowest temperatures is modelled by fits that assume a metallic thermal link to the nuclear stage with a constant heat leak into the noise thermometer (dotted lines).}
\label{fig:5}       
\end{figure}
The current sensing noise thermometers were operated over a wide temperature range from 4.2~K down to 200~$\mathrm{\mu}$K. Over this range there is not a single thermometer against which a comparison can be made. In this work we used a $^3$He melting curve thermometer to determine the Provisional Low Temperature Scale, PLTS-2000 (between 1 and 700~mK); a calibrated germanium thermometer from 50~mK to 4.2~K and a pulsed NMR platinum thermometer, calibrated by the $^3$He melting curve thermometer, was used below 1~mK. Figure \ref{fig:5} shows this comparison for all three thermometers. At each temperature above 6~mK the temperature was stabilised for 1 hour. Excellent agreement between the thermometers was observed over a wide temperature range. Deviations from the reference thermometer are observable up to progressively higher temperature for higher noise sensor resistance. This can be accounted for by a heat leak to the resistor resulting in a temperature gradient across the resistor. Reducing the size of the heat leak through improving the heat sinking and filtering of the wires to the `floating' side of the sensor resistor will be the subject of future development.

\begin{figure}
  \includegraphics[width=0.75\textwidth]{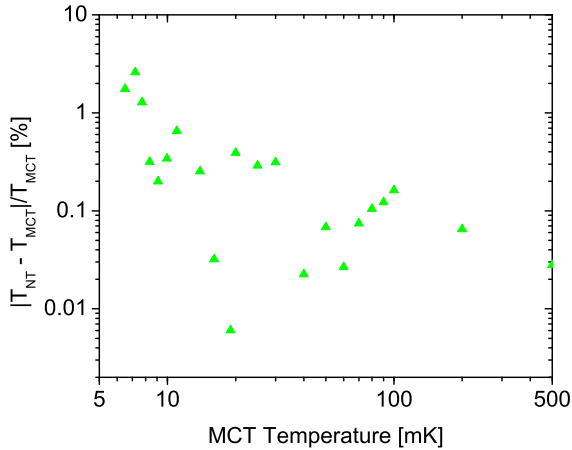}
\caption{Comparison of temperature obtained from the 0.14~$\mathrm{\Omega}$ current sensing noise thermometers with the $^{3}$He melting curve thermometer, down to the lowest temperatures typically reached using a dilution refrigerator. Plotted as a percentage of $\Delta T/T$.}
\label{fig:6}       
\end{figure}

Figure \ref{fig:6} shows the percentage deviation of the comparison between the 0.14~$\mathrm{\Omega}$ sensor and the melting curve thermometer. Here an agreement to better than 1~\% is achieved down to the lowest temperatures attainable on a commercial dilution refrigerator.

It is clear that below 1~mK a relatively low sensor resistance is desirable to ensure good thermalization. The non-contact measurements, which form the basis of the magnetic field fluctuation thermometer \cite{Enss2013} are an alternative method for $T<100\mathrm{\mu}$K. At the lowest temperature obtained, 165~$\mu$K on the platinum thermometer, the 0.2~m$\mathrm{\Omega}$ sensor cooled to 180~$\mu$K.

Another figure of merit for a thermometer is the measurement time required to obtain a given precision of temperature measurement. If we consider capturing a time trace of $N$ points with a time $\Delta t$ between each point, for each point to be an independent measure of the square of the noise current,  $\Delta t$ must be of order the correlation time of the noise or greater (which for this system is the time constant $\tau = L / R$). Therefore in a measuring time $t_{meas}$ one can make approximately $N = t_{meas} / \tau$ independent measurements. A rough estimate for the precision in a given measuring time is thus given by:
\begin{equation}
\label{eq:4}
\frac{\Delta T}{T}\approx \left(\frac{2}{N}\right)^{1/2}=\left(\frac{2 \tau}{t_{meas}}\right)^{1/2}
\end{equation}

The percentage precision in a given measuring time is thus independent of temperature and determined by the noise bandwidth. The noise bandwidth for the sensors used is given in table \ref{tab:1}, and can be seen in figure \ref{fig:3}. Increasing the resistance of the sensor results in an improved precision in a given time, which was the motive behind using PtW resistors with a larger resistance value.

\begin{figure}
  \includegraphics[width=0.75\textwidth]{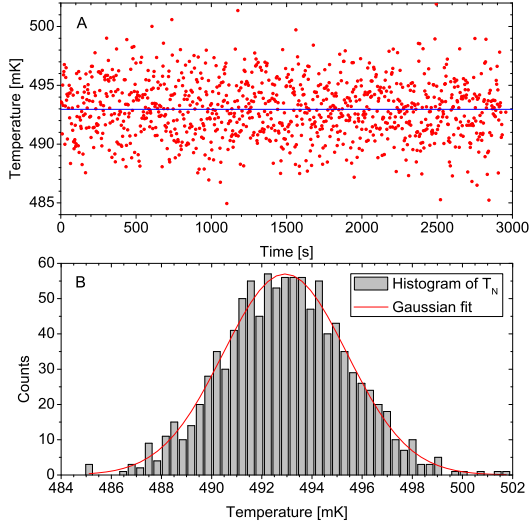}
\caption{(colour online) A: The scatter of temperature measurements using the 1.29~$\mathrm{\Omega}$ PtW current sensing noise thermometer, stabilised with a germanium thermometer at 493~mK. The x axis shows the real time taken to capture the traces. Each temperature reading was acquired in a measurement time of 419~ms. The horizontal line of best fit is the mean temperature. B: A histogram of temperatures showing the scatter follows a normal distribution.}
\label{fig:7}       
\end{figure}

To experimentally determine the precision of the 1.29~$\mathrm{\Omega}$ PtW thermometer, the sample plate was stabilised at a known fixed temperature. Single shot noise traces were taken consecutively. Approximately 1000 individual spectra were captured for accurate statistics. A sampling frequency of 2.5 MS/s with 2$^{20}$ points was used, giving a measurement time of 419 ms per trace. Figure \ref{fig:7}A shows data taken with the PtW thermometer stabilised at 493~mK using the germanium thermometer. This follows a normal distribution as shown by Fig.\ref{fig:7}B, where a Gaussian fit to a histogram of temperatures is shown. The precision is calculated using $\sigma / T$, where $\sigma$ is the standard deviation of the data and $T$ is the mean temperature. In a measurement time of 419 ms, a precision of 0.50 \% was obtained. This precision measurement was repeated at 1.5 K, 100 mK and at 13 mK; the percentage precision was confirmed to be independent of temperature. Reducing the measurement time to 105 ms, resulted in a temperature independent precision of 1.15~\%. We contrast this with the pioneering work on current sensing noise thermometers \cite{Webb1973}, which required a measurement time of 2500~s to achieve a similar performance. The 1.29~$\mathrm{\Omega}$ PtW thermometer is about two orders of magnitude faster than our previous noise thermometers \cite{Lusher2003}, magnetic field fluctuation thermometers \cite{Beyer2007} and on-chip current sensing noise thermometers \cite{Engert2009}, which all have achieved 1 \% precision in times of the order of 10 s. Ultra-low temperature measurements using a contact free cross-correlation noise thermometer \cite{Enss2013} have been reported down to 45~$\mu$K, here using two SQUIDs the cross-correlation technique allows them to reduce the noise contribution from SQUID amplifiers, and the reported measurement time was 400 seconds.

\begin{figure}
  \includegraphics[width=0.75\textwidth]{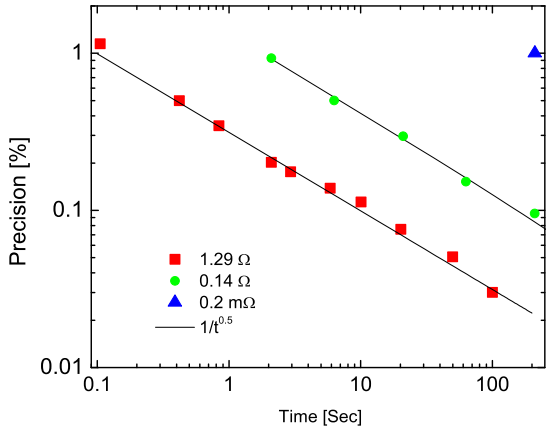}
\caption{Measured precision as a function of measurement time for each of the three thermometers tested, a best fit line proportional to the inverse of the square of the measurement time is shown for the 1.29 and 0.14~$\mathrm{\Omega}$ sensors. }
\label{fig:8}       
\end{figure}

Figure \ref{fig:8} illustrates the dependence of the precision obtained from all three of the current sensing noise thermometers as a function of the measurement time at a temperature of 100~mK. To vary the measurement time, the sampling frequency was kept constant at 50~kS/s, 500~kS/s, 2.5~MS/s for 0.2~m$\mathrm{\Omega}$, 0.14~$\mathrm{\Omega}$ and 1.290~$\mathrm{\Omega}$ respectively; 2$^{20}$ points were captured for each trace and the number of averages was changed to give total measurement times of between 0.1 and 250 seconds. The best fit lines shown are proportional to the inverse of the square root of the measurement time, as in Equation \ref{eq:4}.  The measured precision of 0.03~\% obtained with the 1.29$\mathrm{\Omega}$, sensor in a measurement time of 100~s, is approaching the limit of our ability to stabilise the temperature at 100~mK.

\section{Conclusions}
\label{sec:5}
In conclusion, a series of fast, precise current sensing noise thermometers based on a range of resistive sensors has been tested. Using a single temperature calibration, the noise thermometers can be used to accurately measure a wide range of temperatures. We have presented data on a current sensing noise thermometer that has been optimised for speed, taking advantage of the improvements in SQUID noise and bandwidth. This is two orders of magnitude faster than previous noise thermometers, achieving 1.15~\% precision in just 0.1~seconds. The 0.14~$\mathrm{\Omega}$ sensor tested is in agreement with the $^{3}$He melting curve thermometer to within better than 1~\% over the operating temperature range of a dilution refrigerator, and in this case less than 2 seconds of measurement time is required for 1~\% precision. Finally the lowest value of resistance chosen, 0.2~m$\mathrm{\Omega}$ resulted in a thermometer with a noise temperature of 1~$\mu$K that showed reasonable agreement with a pulsed platinum NMR thermometer to below 200~$\mu$K.

The current sensing noise thermometers described here constitute practical devices for disseminating thermodynamic temperature. The system is user-friendly and could be installed on any commercial ultra-low temperature platform, to provide a reliable and traceable measure of thermodynamic temperature, with high precision (0.1~\%) and high accuracy (better than~1\%) in a reasonable measurement time ($<$~10~s).The potential provision of absolute temperature to all users promises to be a revolutionary technical development in a temperature range expected to be of growing technological and commercial significance.

\begin{acknowledgements}
This work was supported by the European Metrology Research Program (EMRP). The EMRP is jointly funded by the EMRP participating countries within EURAMET and the European Union. Additional support was through the MICROKELVIN project. We acknowledge the support of the European Community - Research Infrastructures under the FP7 Capacities Specific Programme, MICROKELVIN project number 228464.
\end{acknowledgements}

\end{document}